\documentclass[aps,prb,nofootinbib,showpacs,twocolumn]{revtex4}

 \usepackage{amsmath,amssymb,bm,euscript}
 \usepackage{graphicx}
 \usepackage[bf,small]{caption2}
 \usepackage{floatflt}

\def\WRM{Waves Random Media\ }

\def\JPCM{J. Phys.: Condens. Matter\ }
\def\RPP{Rep. Prog. Phys.\ }
\newcommand{\bra}[1]{\ensuremath{\bm{\langle}#1\bm{|}}}
\newcommand{\ket}[1]{\ensuremath{\bm{|}#1\bm{\rangle}}}

\newcommand{\sgn}{\ensuremath\mathrm{\,sgn\,}}

\begin{document}


\title{Spectrum of an open disordered quasi-two-dimensional electron system:\\
       strong orbital effect of the weak in-plane magnetic field}

\author{Yu.\,V. Tarasov}
\email{yutarasov@ire.kharkov.ua}
\affiliation{Institute for Radiophysics and
        Electronics, National Academy of Sciences of Ukraine, 12 Acad.
        Proskura St., Kharkov 61085, Ukraine}

\date{\today}

\begin{abstract}
 The effect of an in-plane magnetic field upon open quasi-two-dimensional
 electron and hole systems is investigated in terms of the carrier
 ground-state spectrum. The magnetic field, classified as weak from the
 viewpoint of correlation between size parameters of classical electron
 motion and the gate potential spatial profile is shown to efficiently cut
 off extended modes from the spectrum and to change singularly the mode
 density of states (MDOS). The reduction in the number of current-carrying
 modes, right up to zero in magnetic fields of moderate strength, can be
 viewed as the cause of magnetic-field-driven metal-to-insulator transition
 widely observed in two-dimensional systems. Both the mode number reduction
 and the MDOS singularity appear to be most pronounced in the mode states
 dephasing associated with their scattering by quenched-disorder potential.
 This sort of dephasing is proven to dominate the dephasing which involves
 solely the magnetic field whatever level of the disorder.
\end{abstract}


\maketitle

\section{Introduction}

The apparent metallic state widely observed in two-dimensional (2D) systems
of Si~MOSFET type as well as in~GaAs/AlGaAs heterostructures
\cite{bib:AbKrSar01,bib:KrSar04} obviously contradicts the well-known
one-parameter scaling theory \cite{bib:AALR79} and as yet has not received
generally accepted theoretical explanation. The existence of such a~state is
mostly believed to result from Coulomb interaction of carriers, which is
rather strong in the systems of low electron and hole density. Estimations of
this interaction indeed can cause the surmise that just this interaction
should lead to quite strong dephasing effect upon electrons which otherwise
would be localized due to scattering by the disorder potential, thus residing
in coherent states. However, at the present time the lack of a~comprehensive
theory for Coulomb interaction in solids precludes from making certain
conclusions about its predominant role in forming the metallic ground state
of 2D electron systems. In particular, a considerable challenge in this
connection is presented by commonplace observations of the dephasing time
saturation in different systems, including 2D ones, at temperatures very
close to zero (see, e.~g., Ref.~\onlinecite{bib:LB02} and numerous references
therein).

Besides the unexpected conduction state of quench-disordered 2D systems, the
metal-insulator transition (MIT) is normally observed there, which currently
has not been proven unambiguously to be determined exclusively by the level
of the disorder. No less puzzling is also the abnormally large response of 2D
electron and hole systems to the relatively weak in-plane magnetic field,
\cite{bib:DKSK92} which is known to significantly suppress the metallic
behaviour of the carriers and even to drive the system into the insulating
regime. \cite{bib:SKSP97,bib:MZVSK01,bib:SKK01,bib:GMRPW02}

In so far as the electrons confined to move in a narrow near-surface
potential well are weakly coupled to the in-plane magnetic field through
their orbital degree of freedom, it is widely believed that such a field
promotes localization of carriers, and thus the MIT, mainly due to strong
spin-related effects. \cite{bib:SHPLRRSG98,bib:OHKY99,bib:Metal99} The
relatively rare papers where orbital coupling was analyzed by taking into
account the finite width of potential wells forming real two-dimensional
systems are not rated by now as fully convincing. Specifically, the
relatively simple model suggested in Ref.~\onlinecite{bib:DSHw00} do not
exhibit sufficiently abrupt transition between metallic and dielectric
regimes, whereas in Ref.~\onlinecite{bib:MFA02} only \emph{corrections}
caused by weak localization of carriers are studied, which can hardly serve
as the conclusive proof for the physical mechanism of the observed effects.

Previously in Refs.~\onlinecite{bib:Tar00},\onlinecite{bib:Tar03},
one-particle theory capable of explaining the metallic ground state as well
as MIT in disordered 2D systems not subjected to magnetic field was developed
starting from basic positions essentially different from those of scaling
theories. Specifically, the conductance of a strictly 2D~\cite{bib:Tar00} and
a~quasi-2D \cite{bib:Tar03} system was calculated in terms of quantum states
pertinent to a perfect finite-size open system of waveguide geometry. In this
approach, the metallic value of the conductance is bound up with the primary
existence of coherent extended waveguide modes rather than one-particle
electronic states originally localized by the disorder. Energy levels of
these \emph{collective} mode states can be widened by quenched disorder
provided that scattering is ensured between \emph{extended} modes having
different \emph{longitudinal} energies. This type of scattering can be viewed
(mathematically) as inelastic, although it is physically provided by a
\emph{static} random potential. In the suggested approach, all extended modes
other than the particular one, if any, can be regarded as the dephasing bath.
With gradual strengthening of the~disorder, the conductance transforms from
its ballistic value in a perfect system, which equals the number of extended
modes times the conductance quantum, to the diffusive value coincident with
standard Drude conductance if the system possesses the number of extended
modes noticeably greater than unity.

In the present work, we apply the mode approach of
Refs.~\onlinecite{bib:Tar00,bib:Tar03} to examine the influence of the
\emph{in-plane} magnetic field upon spectrum of the electrons restricted to
move in a~planar, yet three-dimensional, \emph{open} quantum well. It will be
shown that at large values of the transverse aspect ratio of such an electron
waveguide even a rather weak magnetic field can significantly affect the
electron spectrum. This appears in considerable magnetic-field responsivity
of the number of extended modes, the latter being normally identified as
conducting channels, as well as in MDOS sensitivity to the magnetic field.
The former factor is well known to control the value of the ballistic
conductance of the confined current carriers whereas the latter (MDOS)
governs substantially the mode states dephasing associated with scattering of
the electrons by the disorder potential. Interestingly, mode entanglement
solely due to the magnetic field, with no disorder whatsoever, leads to
electron mass renormalization and does not affect the width of the energy
levels of the collective electron states.

\section{The model}

\begin{figure}[h]
\centering\vspace{\baselineskip}
\scalebox{.5}[.5]{\includegraphics{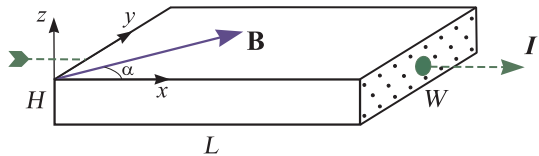}} \caption{The
geometry of Q2D electron waveguide subject to in-plane magnetic
field $\mathbf{B}$.\hfill\label{fig1}}
\end{figure}

Two-dimensional electron and hole systems in practical applications can be
modelled, in view of their open property in the direction of current flow, as
planar quantum waveguides whose transverse structure is governed by lateral
(depletion) potential. The exact form of the confining potential well is of
minor importance for its principal application which reduces to the
restriction of electron transport in the direction normal to the interfacial
area and, consequently, to the ``transverse'' quantization of the electron
spectrum. In this study, we assume the open quasi-two-dimensional (Q2D)
system of carriers having the form of three-dimensional ``electron
waveguide'' of rectangular cross-section, which occupies the coordinate
region
\begin{align}
 & x\in(-L/2,L/2)\ ,\notag\\
 & y\in[-W/2,W/2]\ ,\label{geom}\\
 & z\in[-H/2,H/2] \ ,\notag
\end{align}
as shown in Figure~\ref{fig1}. The length $L$, the width $W$ as well as the
height $H$ of the model system will be regarded as arbitrary except that the
thickness $H$ will be assumed to serve as a small length parameter which will
be specified below. The electron system will be thought of as open at the
ends $x=\pm L/2$ and closed by infinite-wall potentials at all lateral
boundaries. The magnetic field $\mathbf{B}$ is taken to point parallel to
$x$-$y$ plane at arbitrary angle with respect to $x$-axes.

Since main transport coefficients, in particular the conductance, are
expressed in terms of one-particle propagators of carriers, we will analyze
the equation for the retarded Green function of Fermi particles with energy
$\varepsilon_F=k_F^2$. In the Fermi-liquid approximation it has the form
\begin{multline}\label{master_eq}
  \left[\left(\nabla-\frac{2\pi i}{\Phi_0}\mathbf{A}(\mathbf{r})\right)^2+
  k_F^2+i0-V(\mathbf{r})\right]G(\mathbf{r},\mathbf{r}')\\
  =\delta(\mathbf{r}-\mathbf{r}')\ ,
\end{multline}
where $\Phi_0=hc/e$ is the magnetic flux quantum, $\mathbf{A}(\mathbf{r})$ is
the vector potential of the external magnetic field, $V(\mathbf{r})$ is the
scalar random potential due to, say, impurities or rough boundaries of the
confining potential well. We adopt hereinafter the system of units with
$\hbar=2m=1$, $m$ denoting the electron effective mass.

With the magnetic field gauged so as $\mathbf{A}(\mathbf{r})=(B_yz,-B_xz,0)$,
the equation \eqref{master_eq} takes the form
\begin{multline}
  \Bigg[\nabla^2+k_F^2+i0-V(\mathbf{r})-\frac{4\pi i}{\Phi_0}\left(
  B_yz\frac{\partial}{\partial x}-B_xz\frac{\partial}{\partial
  y}\right)\\
  -\left(\frac{2\pi}{\Phi_0}\right)^2
  \mathbf{B}^2z^2\Bigg]G(\mathbf{r},\mathbf{r}')=
  \delta(\mathbf{r}-\mathbf{r}')\ .
  \label{main_eq}
\end{multline}
At this stage it is expedient to go over from the initially three-dimensional
problem to a set of strictly one-dimensional problems, individual for each of
the modes. Towards this end, as a first step one should carry out Fourier
transformation of equation~\eqref{main_eq} over the transverse radius-vector
$\mathbf{r}_{\perp}=(y,z)$. The appropriate set of eigen\-fun\-ctions has the
form
\begin{multline}
  \ket{\mathbf{r}_{\perp};\bm{\mu}}=\frac{2}{\sqrt{WH}}
  \sin\left[\left(\frac{y}{W}+\frac{1}{2}\right)\pi n\right]\\
  \times\sin\left[\left(\frac{z}{H}+\frac{1}{2}\right)\pi m\right] \ ,
  \label{transv_set}
\end{multline}
where $\bm{\mu}=(n,m)$, with $n,m\in\mathbb{N}$, is the vector mode index
conjugate to the coordinate vector $\mathbf{r}_{\perp}$. With functions
\eqref{transv_set}, equation \eqref{main_eq} is readily transformed to the
set of coupled equations for mode Fourier components of the function
$G(\mathbf{r},\mathbf{r}')$, viz.
\begin{multline}\label{mode_eqn}
  \left[\frac{\partial^2}{\partial x^2}+
  k^2_{\bm{\mu}}+i0-
  {\mathcal V}_{\bm{\mu}}(x)\right]G_{\bm{\mu}\bm{\mu}'}(x,x')\\
  -\sum_{\bm{\nu}\neq\bm{\mu}}\hat{{\mathcal U}}_{\bm{\mu}\bm{\nu}}(x)
  G_{\bm{\nu}\bm{\mu}'}(x,x')=
  \delta_{\bm{\mu}\bm{\mu}'}\delta(x-x') \ .
\end{multline}
Here,
\begin{equation}\label{unpert_mode_en}
  k^2_{\bm{\mu}}=k^2_F-\left(\frac{\pi n}{W}\right)^2-\left(\frac{\pi m}{H}\right)^2
\end{equation}
is the unperturbed mode energy,
\begin{equation}\label{V_mu}
  {\mathcal V}_{\bm{\mu}}(x)=V_{\bm{\mu}\bm{\mu}}(x)+
  \frac{H^2}{12l_{B}^4}\left(1-\frac{6}{\pi^2m^2}\right)
\end{equation}
is the diagonal-in-mode-indices matrix element of the total potential which
includes the impurity part $V(\mathbf{r})$ and both of the magnetic terms in
square brackets of equation~\eqref{main_eq},
$l_{B}=\sqrt{\Phi_0/2\pi|\mathbf{B}|}$ is the magnetic length. The term
$V_{\bm{\mu}\bm{\mu}}(x)$ in Eq.~\eqref{V_mu} is the diagonal element of the
mode matrix $\|V_{\bm{\mu}\bm{\nu}}\|$ whose components are evaluated as
\begin{equation}\label{V_mode}
  V_{\bm{\mu}\bm{\nu}}(x)=\int_S\textit{d}\,\mathbf{r}_{\perp}
  \bra{\mathbf{r}_{\perp};\bm{\mu}}V(\mathbf{r})\ket{\mathbf{r}_{\perp};\bm{\nu}}\
  ,
\end{equation}
integration is over cross-section $S$ of the quantum well. Off-diagonal mode
matrix elements $\hat{{\mathcal U}}_{\bm{\mu}\bm{\nu}}(x)$ in
Eq.~\eqref{mode_eqn} also include both the disorder and the magnetic-field
originated potentials, viz.
\begin{equation}\label{U_munu}
  \hat{{\mathcal U}}_{\bm{\mu}\bm{\nu}}(x)=V_{\bm{\mu}\bm{\nu}}(x)+
  2iH\left(\frac{S_{\bm{\mu}\bm{\nu}}^{(x)}}{l_{x}^2W}-
  \frac{S_{\bm{\mu}\bm{\nu}}^{(y)}}{l_{y}^2}\frac{\partial}{\partial x}\right)+
  C_{\bm{\mu}\bm{\nu}}\frac{H^2}{l_{B}^4}\ .
\end{equation}
In Eq.~\eqref{U_munu}, $\ l_i\ $ is the partial magnetic length given by $
l_i^2=\nolinebreak\Phi_0/2\pi|B_i|$, numerical coefficients, which are
specifical for the geometry of the quantum well, have the form
\begin{subequations}\label{SxyC}
\begin{align}
&\begin{aligned} \label{Sx}
 S_{\bm{\mu}\bm{\nu}}^{(x)} =& \varsigma_x\frac{4nn_1}{n^2-n_1^2}
  \sin^2\left[\frac{\pi}{2}(n-n_1)\right]
  \big(1-\delta_{mm_1}\big)\\
 & \times\frac{8mm_1}{\pi^2(m^2-m_1^2)^2}\sin^2\left[\frac{\pi}{2}(m-m_1)\right]\ ,
\end{aligned}\\
&\begin{aligned} \label{Sy}
  S_{\bm{\mu}\bm{\nu}}^{(y)} =& \varsigma_y\delta_{nn_1}
  \big(1-\delta_{mm_1}\big)\\
 & \times\frac{8mm_1}{\pi^2(m^2-m_1^2)^2}
  \sin^2\left[\frac{\pi}{2}(m-m_1)\right]\ ,
\end{aligned}\\
\label{Cpar}
&\begin{aligned} C_{\bm{\mu}\bm{\nu}} =& \delta_{nn_1}
  \big(1-\delta_{mm_1}\big)\\
 & \times\frac{8mm_1}{\pi^2(m^2-m_1^2)^2}
  \cos^2\left[\frac{\pi}{2}(m-m_1)\right]\ ,
\end{aligned}
\end{align}
\end{subequations}
with $\varsigma_i=B_i/|B_i|=\sgn B_i$. In expressions~\eqref{SxyC}, mode
indices are designated such that $\bm{\mu}=(n,m)$ and $\bm{\nu}=\nolinebreak
(n_1,m_1)$.

The potentials ${\mathcal V}_{\bm{\mu}}(x)$ and $\hat{{\mathcal
U}}_{\bm{\mu}\bm{\nu}}(x)$ in Eq.~\eqref{mode_eqn} may be thought of as
responsible for coherent intra-mode and incoherent inter-mode scattering,
respectively. We thus adopt in this work the approach where interactions of
the electron system with both the impurities and the magnetic field are
exploited on equal footing, that is they are treated as the problems of
scattering by additive static potentials which are basically different in
correlation properties only.

\section{Reduction to one-dimensional dynamic problems}

A set of equations \eqref{mode_eqn}, though describing mode propagation in
one spatial dimension, cannot certainly be regarded as a really
one-dimensional dynamical problem by virtue of strong correlation of
different modes via inter-mode potentials \eqref{U_munu}. Formally, this
manifests itself in co-existence in \eqref{mode_eqn} of purely intra-mode
propagators, i.~e. the Green functions having identical mode indices, and
inter-mode propagators $G_{\bm{\nu}\bm{\mu}}(x,x')$ with
$\bm{\nu}\neq\bm{\mu}$.

One can obviate these complications using the method suggested in
Refs.~\onlinecite{bib:Tar00,bib:Tar03}. In the above papers, non-dia\-gonal
elements of the~mode matrix $\|G_{\bm{\mu}\bm{\mu}'}\|$ were proven to be
expressed, by means of some linear operation, through the respective diagonal
elements only. Substitution of thus represented inter-mode propagators into
Eq.~\eqref{mode_eqn} results in the following set of strictly one-dimensional
equations for intra-mode propagators $G_{\bm{\mu}\bm{\mu}}(x,x')$,
\begin{align}
  &\left[\frac{\partial^2}{\partial x^2}+
  k^2_{\bm{\mu}}+i0-
  {\mathcal V}_{\bm{\mu}}(x)-\hat{\mathcal T}_{\bm{\mu}}\right]
  G_{\bm{\mu}\bm{\mu}}(x,x')=\delta(x-x')\ ,\notag\\
  &\quad\text{for}\ \forall\bm{\mu}\ .
\label{G_mode-diag}
\end{align}
Here, $\hat{\mathcal T}_{\bm{\mu}}$ is the \emph{operator} (integral)
potential, well-known as the $T$-matrix in the quantum theory of
scattering,\cite{bib:Newton68,bib:Taylor72} which acts in the $x$-coordinate
space~$\mathbb{X}$. It has the form
\begin{equation}\label{T-oper}
  \hat{\mathcal T}_{\bm{\mu}}=\bm{P}_{\bm{\mu}}\hat{\mathcal U}
  (\openone-\hat{\mathsf R})^{-1}\hat{\mathsf R}\bm{P}_{\bm{\mu}}\ ,
\end{equation}
where $\hat{\mathcal U}$ and $\hat{\mathsf R}$ are the operators acting in
the mixed mode-coordinate space ${\mathsf{\overline M}_{\bm{\mu}}}$
constructed as a direct pro\-duct of the space $\mathbb{X}$ and the truncated
mode space which incorporates the whole set of mode indices except the unique
mode index $\bm{\mu}$. The operator $\hat{\mathcal U}$ is specified in
${\mathsf{\overline M}_{\bm{\mu}}}$ by its matrix elements
\begin{equation}\label{U-oper}
  \bra{x,\bm{\nu}}\hat{\mathcal U}\ket{x',\bm{\nu'}} =
  {\mathcal U}_{\bm{\nu}\bm{\nu'}}(x)\delta(x-x')\ ,
\end{equation}
matrix elements of the operator $\hat{\mathsf R}$ have the form
\begin{equation}\label{R_matr_el}
  \bra{x,\bm{\nu}}\hat{\mathsf R}\ket{x',\bm{\nu'}} =
  G^{(V)}_{\bm{\nu}}(x,x')\hat{{\mathcal U}}_{\bm{\nu}\bm{\nu'}}(x')\ .
\end{equation}
The function $G^{(V)}_{\bm{\nu}}(x,x')$ in \eqref{R_matr_el} will be thought
of as the \emph{trial} mode Green function which satisfies the equation
resulting from Eq.~\eqref{mode_eqn} provided that all inter-mode potentials
are put identically equal to zero,
\begin{equation}\label{G_trial}
  \left[\frac{\partial^2}{\partial x^2}+
  k^2_{\bm{\mu}}+i0-{\mathcal V}_{\bm{\mu}}(x)\right]G^{(V)}_{\bm{\mu}}(x,x')=
  \delta(x-x')\ .
\end{equation}
The operator $\bm{P}_{\bm{\mu}}$ in \eqref{T-oper} is the projection operator
whose action reduces to assigning the given value $\bm\mu$ to the nearest
mode index of an arbitrary operator standing next to it (either to the left
or right), without affecting the product in the $\mathbb{X}$ space.

With intra-mode Green functions found from the set of
equations~\eqref{G_mode-diag}, all inter-mode propagators are expressed via
the operator relation
\begin{equation}\label{Inter<->Intra}
  \hat{G}_{\bm{\nu}\bm{\mu}}=\bm{P}_{\bm{\nu}}(\openone-\hat{\mathsf R})^{-1}\hat{\mathsf R}
  \bm{P}_{\bm{\mu}}\hat{G}_{\bm{\mu}\bm{\mu}}\ ,
\end{equation}
$\hat{G}_{\bm{\nu}\bm{\mu}}$ and $\hat{G}_{\bm{\mu}\bm{\mu}}$ being thought
of as matrices in $\mathbb{X}$-space. The initially three-dimensional problem
thus reduces to the~set of separate 1D equations \eqref{G_mode-diag}, each
representing the closed problem provided that trial Green functions are
independently found from equation~\eqref{G_trial}.

To analyze the mode states spectrum, i.~e. the spectrum of differential
operator in Eq.~\eqref{G_mode-diag}, it is worthwhile to renormalize mode
energies in Eqs.~\eqref{G_mode-diag} and \eqref{G_trial} by extracting from
the initial mode energy \eqref{unpert_mode_en} the non-random ``magnetic''
part of the intra-mode potential ${\mathcal V}_{\bm{\mu}}(x)$, thus defining
the new ``unperturbed'' mode energy
\begin{equation}\label{moden-renorm}
  \varkappa_{\bm{\mu}}^2=k^2_{\bm{\mu}}-
  \frac{H^2}{12l_{B}^4}\left(1-\frac{6}{\pi^2m^2}\right)\ .
\end{equation}
In such a way one is led to solve, in place of equations \eqref{G_mode-diag}
and \eqref{G_trial}, a couple of different, though equivalent, equations,
namely
\begin{subequations}\label{Eqs-renorm}
\begin{align}\label{Eqs-renorm-main}
  &\begin{aligned}
  \left[\frac{\partial^2}{\partial x^2}+
  \varkappa^2_{\bm{\mu}}+i0-
  V_{\bm{\mu}\bm{\mu}}(x)-\hat{\mathcal T}_{\bm{\mu}}\right]
  G&_{\bm{\mu}\bm{\mu}}(x,x')\\&=\delta(x-x')
  \end{aligned}
  \\\intertext{and}
  & \left[\frac{\partial^2}{\partial x^2}+
  \varkappa^2_{\bm{\mu}}+i0-V_{\bm{\mu}\bm{\mu}}(x)\right]G^{(V)}_{\bm{\mu}}(x,x')=
  \delta(x-x')\ ,
  \label{Eqs-renorm-trial}
\end{align}
\end{subequations}
which must be supplied with correct boundary conditions at open ends of the
system, to be discussed in the next section.

\section{Spectrum of the mode states}

To examine differential operator \eqref{Eqs-renorm-main}, which is of
principal import for our purpose, one should first solve
Eq.~\eqref{Eqs-renorm-trial} for the truly one-dimensional trial Green
function. In the absence of magnetic field, this problem was resolved in
Ref.~\onlinecite{bib:Tar00} where arbitrary statistical moments of the
function $G^{(V)}_{\bm{\mu}}(x,x')$ were found in the case of open system
subject to weak disorder potential by applying the averaging technique
appropriate for causal-type random functionals. The condition for weak
impurity scattering (WIS), which we assume to hold in this study as well, can
be cast to the form of the inequality pair
\begin{equation}\label{weak_imp}
  k^{-1}_F,\,r_c\ll\ell\ ,
\end{equation}
were $r_c$ is the correlation radius of the random potential, $\ell$ is the
electron mean free path relating to it.

In the case of non-zero magnetic field the solution to
Eq.~\eqref{Eqs-renorm-trial} is much more involved than that accomplished in
Ref.~\onlinecite{bib:Tar00}. This will be thoroughly examined in a separate
publication, while here we outline the solution along with criteria of its
applicability.

Given the magnetic potentials, to adequately take into account the open
property of one-dimensional system governed by equation
\eqref{Eqs-renorm-trial} one should explore this equation on the extended
$x$-axis rather than on dis\-ordered and subject to magnetic field interval
$\mathcal{L}=\nolinebreak(-L/2,L/2)$. Being considered on the whole axis,
equation \eqref{Eqs-renorm-trial} describes the motion of a quantum particle
created with energy $k^2_{\bm{\mu}}$ at point $x'$ and then propagating in
two-component scalar potential with bounded support. The regular component of
this combined potential is due to the magnetic field. It has the form
$V^{(reg)}(x)=\nolinebreak\theta(L/2-\nolinebreak|x|)\frac{H^2}{12l_{B}^4}
\left(1-\frac{6}{\pi^2m^2}\right)$, whereas the random component, which is of
impurity origin, is covered by the function
$V^{(ran)}(x)=\theta(L/2-\nolinebreak|x|)V_{\bm{\mu}\bm{\mu}}(x)$.

To perform configurational averaging over the random part of the potential it
is worthwhile to express the trial Green function in terms of wave functions
of \emph{causal} type rather than functions that meet the initially stated
boundary-value (BV) problem. This may be achieved by employing the formula
(for the sake of clarity we omit mode index $\bm{\mu}$)
\begin{align}
  G^{(V)}(x,x')=\frac{1}{\mathcal{W}}
  \big[\psi_+&(x)\psi_-(x')\theta(x-x')\notag\\
  &+\psi_+(x')\psi_-(x)\theta(x'-x) \big] \ ,
\label{Green-Cochi}
\end{align}
where $\psi_{\pm}(x)$ are two different solutions of homogeneous equation
\eqref{G_trial} with boundary conditions specified for each of them at only
one end of the system, viz. $x\to\pm\infty$, depending on the sign index,
$\mathcal{W}$~is the Wronskian of those solutions. With this representation,
the trial propagator itself meets, as it must, the initial BV problem.

The openness of the finite-size system under consideration implies that far
from the source coordinate~$x'$, specifically at $x\to\pm\infty$, Green
function $G^{(V)}(x,x')$ must have the~form of \emph{outgoing} free waves. In
view of boundedness of the support of the potentials, at large values
of~$|x|$ functions $\psi_{\pm}(x)$ have to be taken as
\begin{equation}\label{psi_rad}
  \psi_{\pm}(x)=c_{\pm}\exp\big[\pm ik(x-L/2)\big]\ .
\end{equation}

Inside the magnetically biased interval $\mathcal{L}$, in order to properly
take into account the electron backscattering from the potential
$V^{(ran)}(x)$, it is worthwhile to seek wave functions $\psi_{\pm}(x)$ in
the form
\begin{align}\label{psi_pm-in}
  \psi_{\pm}(x)=\pi_{\pm}(x)\mathrm{e}^{i\varkappa (\pm x - L/2)}-
  i\gamma_{\pm}(x)\mathrm{e}^{-i\varkappa (\pm x - L/2)}\ ,
\end{align}
with $\varkappa^2$ specified in \eqref{moden-renorm}. Under WIS conditions
\eqref{weak_imp}, envelope functions $\pi_{\pm}(x)$ and $\gamma_{\pm}(x)$ in
\eqref{psi_pm-in} can be regarded as smooth factors in comparison with
near-standing fast exponentials, which leads to the following coupled dynamic
equations,
\begin{subequations}\label{dyn-eq_pi-gamma}
\begin{align}
 \label{dyn-eq_pi-gamma-1}
  & \pm\pi'_{\pm}(x)+
  i\eta(x)\pi_{\pm}(x)+\zeta^*_{\pm}(x)\gamma_{\pm}(x)=0\ ,\\
 \label{dyn-eq_pi-gamma-2}
  & \pm\gamma'_{\pm}(x)-
  i\eta(x)\gamma_{\pm}(x)+\zeta_{\pm}(x)\pi_{\pm}(x)=0\ .
\end{align}
\end{subequations}
Random functions $\eta(x)$ and $\zeta_{\pm}(x)$ in
Eqs.~\eqref{dyn-eq_pi-gamma} are constructed as normalized packets of spatial
harmonics of the impurity potential ~$V(x)\equiv V_{\bm{\mu}\bm{\mu}}(x)$,
which have the form
\begin{subequations}\label{eff_fields}
\begin{align}
 \label{eff_fields-eta}
  & \eta(x)=\frac{1}{2\varkappa}\int_{x-l}^{x+l}\frac{\mathrm{d}t}{2l}
  V(t)\ ,\\
 \label{eff_fields-zeta}
  & \zeta_{\pm}(x)=\frac{1}{2\varkappa}\int_{x-l}^{x+l}\frac{\mathrm{d}t}{2l}
  V(t)\exp[2i\varkappa(\pm x-L/2)]\ .
\end{align}
\end{subequations}
Spatial averaging in \eqref{eff_fields} is carried out over the interval $2l$
of arbitrary length intermediate between small lengths $\varkappa^{-1}$ and
$r_c$, on the one hand, and the large scattering length (to be determined
self-consistently), on the other. In view of these limitations, the
``potentials'' $\eta(x)$ and $\zeta_{\pm}(x)$ provide forward and backward
scattering of harmonics $\pm\varkappa$, respectively.

By joining the solutions \eqref{psi_pm-in} and \eqref{psi_rad} at the end
points of the interval $\mathcal{L}$ we arrive at the exact boundary
conditions for the envelopes $\pi_{\pm}(x)$ and $\gamma_{\pm}(x)$, viz.
\begin{subequations}\label{bound_cond}
\begin{align} \label{bc-pi}
  \pi_{\pm}(\pm L/2)= & \mathrm{const}\ ,\\
  \label{bc-gamma}
  \gamma_{\pm}(\pm L/2)= &
  \EuScript{R}^{(B)}\pi_{\pm}(\pm L/2)\ .
\end{align}
\end{subequations}
The quantity
\begin{equation}\label{magn_refl}
  \EuScript{R}^{(B)}=-i\frac{k-\varkappa}{k+\varkappa}\ ,
\end{equation}
as it follows from \eqref{psi_pm-in}, is the amplitude reflection coefficient
from the boundary between magnetically biased and unbiased regions. This
reflection will be hereinafter referred to as ``magnetic scattering''
associated with the above introduced potential $V^{(reg)}(x)$.

Below in this paper, scattering associated with both of the potentials,
$V^{(ran)}(x)$ and $V^{(reg)}(x)$, will be regarded as weak. The weakness of
the impurity scattering implies the inequalities \eqref{weak_imp} whereas the
magnetic scattering will be thought of as weak provided that the requirement
is met $|\EuScript{R}^{(B)}|\ll 1$. From \eqref{bc-gamma} and
\eqref{moden-renorm} one can make sure that in terms of appropriate physical
parameters the condition for weak magnetic scattering (WMS) may be expressed
as the inequality
\begin{equation}\label{WMS-cond}
  \left(\frac{H}{R_c}\right)^2\ll 1\ ,
\end{equation}
where $R_c=k_Fl_B^2$ is the maximal classical cyclotron radius of the
electron orbit. It is just from the viewpoint of this constraint that we will
regard the in-plane magnetic field to be weak.

As far as the impurity scattering is concerned, to do the averaging over
realizations of the potential $V(\mathbf{r})$ this random function will be
thought of to possess the following correlation properties,
\begin{subequations}\label{imp_corr}
\begin{align}
 \label{corr1}
  \big<V(\mathbf{r})\big>&=0\ ,\\
 \label{corr3d_simpl}
  \big<V(\mathbf{r})V(\mathbf{r}')\big>&=\mathcal{QW}(x-x')
  \delta(\mathbf{r}_{\perp}-\mathbf{r}'_{\perp}) \ ,
\end{align}
\end{subequations}
angular brackets denote configurational averaging. Under WIS conditions
\eqref{weak_imp}, the equalities \eqref{imp_corr} are sufficient to
adequately accomplish the averaging for rather wide class of the random
potential statistics, since in this case function $V(\mathbf{r})$  may be
regarded as approximately Gaussian distributed.\cite{bib:LGP82}

By applying the averaging technique outlined in the Appendix the average
trial Green function is obtained in the following form, which is valid
provided WIS and WMS conditions hold simultaneously,
\begin{widetext}
\begin{equation}\label{G^V_nu}
  \left<G^{(V)}_{\bm{\mu}}(x,x')\right>\approx\frac{-i}{2\varkappa_{\bm{\mu}}}
  \exp\left\{\left[i\varkappa_{\bm{\mu}}-\frac{1}{2}\left(\frac{1}{L_f^{(V)}(\bm{\mu})}+
  \frac{1}{L_b^{(V)}(\bm{\mu})}\right)\right]|x-x'|\right\}\ .
\end{equation}
\end{widetext}
Here,
\begin{subequations}\label{ext_length}
\begin{align}
\label{ext_length-Lf}
  &L_f^{(V)}(\bm{\mu})=\left(\frac{4}{3}\right)^2\frac{S}{\mathcal{Q}}
  \varkappa^2\\
\intertext{and}
  &L_b^{(V)}(\bm{\mu})=
  \frac{L_f^{(V)}(\bm{\mu})}%
  {\widetilde{\mathcal{W}}(2\varkappa_{\bm{\mu}})}
\label{ext_length-Lb}
\end{align}
\end{subequations}
are the extinction lengths related to forward ($f$) and backward ($b$)
scattering by the potential $V_{\bm{\mu}\bm{\mu}}(x)$,
$\widetilde{\mathcal{W}}(q)$ is the Fourier transform of function
$\mathcal{W}(x)$ from \eqref{corr3d_simpl}. With the result \eqref{G^V_nu},
the average square norm of the operator $\hat{\mathsf R}$ from \eqref{T-oper}
can be represented as a sum of ``impurity'' and ``magnetic'' terms, viz.
$\big<\|\hat{\mathsf R}\|^2\big>\approx \big<\|\hat{\mathsf
R}^{(imp)}\|^2\big>+\big<\|\hat{\mathsf R}^{(B)}\|^2\big>$, which are
estimated as
\begin{subequations}\label{Rnorm_est}
\begin{align}\label{Rnorm_est-imp}
  &\big<\|\hat{\mathsf R}^{(imp)}\|^2\big> \sim \frac{1}{k_F\ell}\ll 1\ , \\
\label{Rnorm_est-B}
  &\big<\|\hat{\mathsf R}^{(B)}\|^2\big> \sim \left(\frac{H}{R_c}\right)^2\ll 1\ .
\end{align}
\end{subequations}
Inequalities \eqref{Rnorm_est} permit simplification of the operator
potential $\hat{\mathcal T}_{\bm{\mu}}$ since the inverse operator in
\eqref{T-oper} can be approximately replaced with the unit operator. The
inter-mode potential thus reduces to the relatively simple form,
\begin{equation}\label{T-approx}
  \hat{\mathcal T}_{\bm{\mu}}\approx
  \bm{P}_{\bm{\mu}}\hat{\mathcal U}
  \hat{\mathcal G}^{(V)}\hat{\mathcal U}\bm{P}_{\bm{\mu}} \ ,
\end{equation}
where $\hat{\mathcal G}^{(V)}$ stands for the operator in ${\mathsf{\overline
M}_{\bm{\mu}}}$ which is specified by the matrix elements of the following
form,
\begin{equation}\label{GV-matr}
  \bra{x,\bm{\nu}}\hat{\mathcal G}^{(V)}\ket{x',\bm{\nu}'}=
  G^{(V)}_{\bm{\nu}}(x,x')\delta_{\bm{\nu}\bm{\nu}'}\ .
\end{equation}

Unlike quasi-local intra-mode potential $V_{\bm{\mu}\bm{\mu}}(x)$, the
operator potential \eqref{T-approx} possesses, even in the absence of
magnetic field, the nonzero mean value. Therefore, to apply further a
perturbation theory it is reasonable to represent this operator as a sum of
averaged and fluctuating parts, i.~e. $\hat{\mathcal
T}_{\bm{\mu}}=\big<\hat{\mathcal
T}_{\bm{\mu}}\big>+\nolinebreak\Delta\hat{\mathcal T}_{\bm{\mu}}$. With
regard to Eq.~\eqref{U_munu}, the mean operator $\big<\hat{\mathcal
T}_{\bm{\mu}}\big>$ can be splitted (though quite conventionally) into the
local ``impurity'' and essentially non-local ``magnetic'' terms. The action
of short-correlated impurity part of this operator reduces to multiplication
of the mode propagator by the complex self-energy
factor,\cite{bib:Tar00,bib:Tar03}
\begin{equation}
\Big[\big<\hat{\mathcal
T}^{(imp)}_{\bm{\mu}}\big>G_{\bm{\mu}\bm{\mu}}\Big](x,x')
=-\varSigma^{(imp)}_{\bm{\mu}}G_{\bm{\mu}\bm{\mu}}(x,x')\ ,
\end{equation}
where $\varSigma^{(imp)}_{\bm{\mu}}=\Delta\varkappa_{\bm{\mu}}^2+
i/\tau_{\bm{\mu}}^{(\varphi)}$ and the notations are used
\begin{subequations}\label{renorm_imp}
\vspace{-.5cm}%
\begin{align}
\label{ren_en}
  \Delta \varkappa_{\bm{\mu}}^2 &=
  \frac{\mathcal{Q}}{S}\sum_{\bm{\nu}\neq\bm{\mu}}\mathcal{P}
  \int_{-\infty}^{\infty}\frac{\mathrm{d}q}{2\pi} \;
  \frac{\widetilde{\mathcal{W}}(q+\varkappa_{\bm{\mu}})}{q^2-\varkappa_{\bm{\nu}}^2}
  \ , \\
\label{dephase}
  \frac{1}{\tau_{\bm{\mu}}^{(\varphi)}} &=
  \frac{\mathcal{Q}}{4S}
  \overline{\sum_{\bm{\nu}\neq\bm{\mu}}} \frac{1}{\varkappa_{\bm{\nu}}}
  \left[\widetilde{\mathcal{W}}(\varkappa_{\bm{\mu}}-\varkappa_{\bm{\nu}})+
  \widetilde{\mathcal{W}}(\varkappa_{\bm{\mu}}+\varkappa_{\bm{\nu}})\right]
  \ .
\end{align}
\end{subequations}
Symbol $\mathcal{P}$ in \eqref{ren_en} stands for the integral principal
value, the bar over the summation index in \eqref{dephase} signifies that the
summation is carried out over extended modes only. The conditional character
of the term ``impurity self energy'' with reference to
$\varSigma^{(imp)}_{\bm{\mu}}$ is related to the mere fact that this factor
is actually determined by both the impurity potential, whose correlator is
proportional to the factor of $\mathcal{Q}$, and the magnetic field, which
renormalizes the wavenumbers $\varkappa_{\bm{\mu},\bm{\nu}}$ and also adjusts
the number of extended modes, see next subsection.

The action of the expressly non-local ``magnetic'' part of the operator
$\big<\hat{\mathcal T}_{\bm{\mu}}\big>$ is specified by the formula
\begin{widetext}
\vspace{-.4cm}
\begin{eqnarray}
  \big[\big<\hat{\mathcal
  T}_{\bm{\mu}}^{(B)}\big>G_{\bm{\mu}\bm{\mu}}\big](x,x')&=&
  \sum_{\bm{\nu}\neq\bm{\mu}}
  \left[2iH\left(\frac{S_{\bm{\mu}\bm{\nu}}^{(x)}}{Wl_{x}^2}-
  \frac{S_{\bm{\mu}\bm{\nu}}^{(y)}}{l_{y}^2}\frac{\partial}{\partial
  x}\right)+ C_{\bm{\mu}\bm{\nu}}\frac{H^2}{l_{B}^4}\right]
  \notag\\
 \label{TBav}
  && \times\int_L\textit{d}x_1\left<G^{(V)}_{\bm{\nu}}(x,x_1)\right>
  \left[2iH\left(\frac{S_{\bm{\bm{\nu}\mu}}^{(x)}}{Wl_{x}^2}-
  \frac{S_{\bm{\nu}\bm{\mu}}^{(y)}}{l_{y}^2}\frac{\partial}{\partial
  x_1}\right)+ C_{\bm{\nu}\bm{\mu}}\frac{H^2}{l_{B}^4}\right]
  G_{\bm{\mu}\bm{\mu}}(x_1,x')\ ,
\end{eqnarray}
\end{widetext}
\vspace{-.2cm}%
wherefrom the ``magnetic'' self-energy, which is applicable under WS
conditions, is immediately deduced
\begin{align}\label{B_selfenergy}
  \varSigma^{(B)}_{\bm{\mu}}=
  -4H^2\sum_{\bm{\nu}\neq\bm{\mu}}
  \frac{\left(\frac{S_{\bm{\mu}\bm{\nu}}^{(x)}}{Wl_{x}^2}-
  i\varkappa_{\bm{\mu}}\frac{S_{\bm{\mu}\bm{\nu}}^{(y)}}{l_{y}^2}\right)
  \left(\frac{S_{\bm{\nu}\bm{\mu}}^{(x)}}{Wl_{x}^2}-
  i\varkappa_{\bm{\mu}}\frac{S_{\bm{\nu}\bm{\mu}}^{(y)}}{l_{y}^2}\right)}
  {\varkappa_{\bm{\mu}}^2-\varkappa_{\bm{\nu}}^2-
  i\varkappa_{\bm{\nu}}\left(\frac{1}{L_f^{(V)}(\bm{\nu})}+
  \frac{1}{L_b^{(V)}(\bm{\nu})}\right)}
  \ .
\end{align}

\subsection{Mode content of the open quantum system}

Both the impurity and the magnetic self-energies are complex-valued
quantities, whose real parts renormalize mode energies whereas the imaginary
parts determine the uncertainty of energy levels. The requirement for mode
energies to be positive defined specifies the number of extended modes in
the~quantum system, which is normally referred to as the number of conducting
channels,~$N_c$. Computation of exact number of these modes, though clear in
principle, is an intricate problem in general. For the system under
consideration the number $N_c$ can be most easily found in the particular
case of the magnetic field oriented lengthwise with respect to the current
direction, i.~e. for $\mathbf{B}\parallel Ox$. In this case mode energy
renormalization due to the \emph{inter-mode} magnetic scattering, which is
covered by the magnetic self-energy~\eqref{B_selfenergy}, is small as
compared with the \emph{intra-mode} magnetic correction present in the mode
energy \eqref{moden-renorm}. Taking account of this fact, one can calculate
the number of extended modes as
\begin{subequations}\label{Nc-def}
\vspace{-.4cm}
\begin{equation}\label{Nc=sum}
  N_c=\sum_{m=1}^{N_c^{(z)}}N_c^{(y)}(m)\ ,
\vspace{-.3cm}
\end{equation}
where
\vspace{-.2cm}
\begin{equation}\label{Nc(z)}
  N_c^{(z)}=\mathrm{int}\left[\frac{k_FH}{\pi}
  \sqrt{1-\frac{H^2}{12R_c^2}}\right]
\vspace{-.1cm}
\end{equation}
is the number of quantization levels in $z$-direction, whose energies lie
beneath the Fermi energy, and
\begin{widetext}
\vspace{-.2cm}
\begin{equation}\label{Nc(y)}
  N_c^{(y)}(m)=\mathrm{int}\left[\frac{k_FW}{\pi}
  \sqrt{1-\left(\frac{\pi m}{k_FH}\right)^2-
  \frac{H^2}{12R_c^2}\left(1-\frac{6}{\pi^2m^2}\right)}\right]
\end{equation}
\end{widetext}
\end{subequations}
is the number of $y$-directional quantization levels pertinent to the $m$-th
level of $z$-quantization. Symbol $\mathrm{int[\ldots]}$ in \eqref{Nc(z)} and
\eqref{Nc(y)} denotes the integer part of the number enclosed in square
brackets.

The sum \eqref{Nc=sum} can be readily evaluated in the case where the number
of extended modes relating to both of the transverse axes of the quantum
waveguide is large as compared to unity. By replacing the sum with the
integral one readily gets
\begin{equation}\label{Nc-magn}
  N_c\approx \frac{k_F^2S}{4\pi}\left(1-\frac{H^2}{12R_c^2}\right)\ ,
\end{equation}
wherefrom it is evident that application of the in-plane magnetic field can
significantly reduce the number of extended modes, even though inequality
\eqref{WMS-cond} holds true. This reduction is definitely the geometrical
effect which is due to the curving of the electron orbits in the magnetic
field, and thus it can be only taken into consideration within the model of a
finite-width quantum well that forms a 2D system.

\begin{figure}[h]
\centering \setcaptionmargin{1em}
{\includegraphics[width=.9\columnwidth]{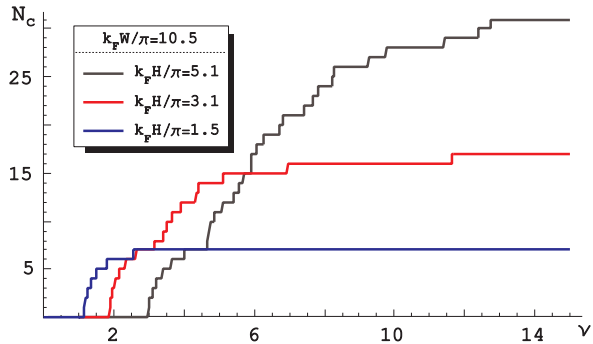}}
\caption{The magnetic-field dependence of the number of effective conducting
channels. The parameter $\nu$ is the Landau filling factor,
$\nu=(k_Fl_B)^2$.\hfill\label{fig2}}
\end{figure}

In Fig.~\ref{fig2}, the numerical results for the number of effective
conducting channels as a function of the inverse magnetic field scaled
as the Landau filling factor $\nu=\nolinebreak (k_Fl_B)^2=\nolinebreak
k_FR_c$ are presented. The collapse of the number of current-carrying modes
with a growth in the magnetic field is apparent, regardless of the quantum
waveguide thickness~$H$, the width $W$ is assumed constant. The in-plane
rotation of the magnetic field smoothly changes the presented picture because
the real part of self-energy \eqref{B_selfenergy} can at most reach the same
(on the order of magnitude) value as the intra-mode magnetic addend in
\eqref{moden-renorm}.\cite{rem1}

\begin{figure}[h]
\centering
\setcaptionmargin{1em}
{\includegraphics[width=.95\columnwidth]{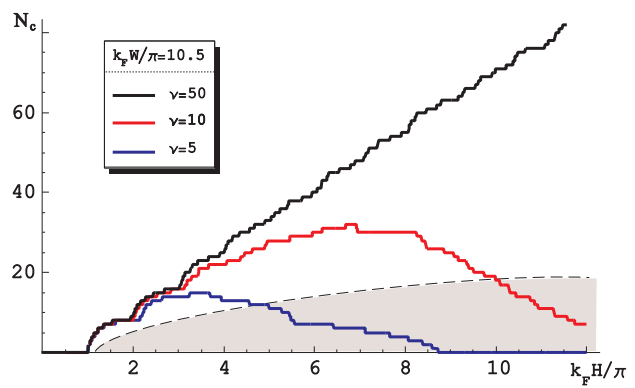}} \caption{The
number of conducting channels vs the width of the near-surface
potential well at different values of the Landau filling factor.
The shaded area below the dashed curve covers the parameter region
where the criterion \eqref{WMS-cond} is violated.
\hfill\label{fig3}}
\end{figure}

In Fig.~\ref{fig3}, the relation between the number of channels and the
effective thickness of the quantum waveguide is presented, which actually
demonstrates the dependence of $N_c$ on the depletion voltage adjusting the
width of the near-surface potential well. In the extremely low magnetic field
(black curve) the number of channels increases nearly linear with growing
$H$, in accordance with standard geometrical consideration applicable to
systems of waveguide configuration, and also with the conventional Ohm's law
which is undoubtedly valid for bulk conductors. With the growing magnetic
field, the
conventional geometric increase in the number of channels gets slower,
gradually indicating the trend for lowering the number of conducting modes.
This unusual dependence of the mode content of the electron waveguide on its
effective thickness is due to non-monotonic dependence on $H$ of the mode
energy~\eqref{moden-renorm}.

Obviously, on a further increase of the magnetic field the tendency towards
lowering the number of conducting channels must be stabilized owing to terms
$\propto l_B^{-4}$ in square brackets in r.h.s. of Eq.~\eqref{TBav}. However,
this can happen only in the domain of relatively strong magnetic fields,
where the WMS condition is violated and the approximate expression
\eqref{T-approx} for the inter-mode potential is no longer applicable. In
such magnetic fields, the bulk quantum Hall effect is expected to come in the
foreground, which is beyond the scope of this paper.

\subsection{Dephasing of the mode states: the magnetic-field driven disorder}

Besides the impact on the number of extended quantum modes whose transverse
energies are beneath the Fermi level, the in-plane magnetic field can
signi\-fi\-cantly affect the coherent properties of the conducting channels.
This field controls the imaginary parts of both the impurity-governed
self-energy \eqref{renorm_imp} and the magnetic
self-energy~\eqref{B_selfenergy}. Both of these self-energies arise due to
the inter-mode scattering. One should bear in mind, however, that
$\varSigma^{(imp)}_{\bm{\mu}}$ is basically determined by scattering from the
impurity potential whereas the magnetic self-energy,
$\varSigma^{(B)}_{\bm{\mu}}$, originates in the main from mode mixing due to
the orbital effect of in-plane magnetic field.

It is important to note that intermixing of channels which is controlled
solely by the magnetic field cannot result in significant dephasing of mode
states. By comparing the imaginary part of self-energy \eqref{B_selfenergy}
and the level width \eqref{dephase} one can determine that the ratio of
``purely magnetic'' and ``impurity-governed'' dephasing rates is evaluated as
\begin{equation}\label{B/imp}
  \frac{\Im\varSigma^{(B)}_{\bm{\mu}}}{\Im\varSigma^{(imp)}_{\bm{\mu}}}\sim
  \left[\frac{H}{R_ck_FW}\left(\frac{B_x}{B}\right)^2+
  \frac{H}{R_c}\left(\frac{B_y}{B}\right)^2\right]^2\ll 1\ .
\end{equation}
This implies that under WMS condition the magnetic-field originated dephasing
is negligible, what\-ever strength of the disorder. The conclusion is thus
unavoidable that strong inter-mode mixing resulting from the magnetic field
cannot give rise to significantly widening the mode levels unless there
exists some \emph{random} potential due to, say, impurities or the roughness
of quantum well boundaries, which can mediate the dephasing effect of the
magnetic field. The specific role of the magnetic field, as far as the mode
entanglement is concerned, reduces to the change in collective parameters of
the electron motion, such as the mode content of the confined system and the
mode density of states, and in such an indirect way to modification of
\emph{scattering parameters} pertinent to random generators of inter-mode
transitions (i.~e., the impurity scattering cross-section, the polar pattern
of electron reflection from rough boundaries, etc.).

The influence of the magnetic field upon transport parameters manifests
itself directly through the mode dephasing rate. Analytically, the estimate
of this quantity can be most easily deduced from Eq.~\eqref{dephase} in the
case where the number of quantization levels related to both of the
transverse directions is large as compared to unity and the sum in
Eq.~\eqref{dephase} can be replaced with the integral. The dephasing rate for
the particular mode $\bm{\mu}$ in this case reads
\begin{equation}\label{deph(B)}
  \frac{1}{\tau_{\bm{\mu}}^{(\varphi)}(B)}\approx
  \frac{1}{\tau_{\bm{\mu}}^{(\varphi)}(0)}\sqrt{1-
  \frac{H^2}{12R_c^2}}\ ,
\end{equation}
where $1/\tau_{\bm{\mu}}^{(\varphi)}(0)=k_F\mathcal{Q}/4\pi$ is the
$\bm{\mu}$-th mode level width attributed to scattering due to the disorder
potential only, with no external magnetic field.\cite{bib:Tar03} The value of
this zero-field level width equals exactly half the inverse mean free time
calculated within the framework of classical kinetic theory. Note that in the
domain of weak magnetic fields corresponding to inequality \eqref{WMS-cond}
the dephasing rate \eqref{deph(B)} decreases nearly quadratically in the
magnetic field and has universal value, the same for each of the extended
modes.\cite{bib:Tar03}

\begin{figure}[h]
\vspace{.5\baselineskip}\centering\setcaptionmargin{1em}
{\includegraphics[width=.95\columnwidth]{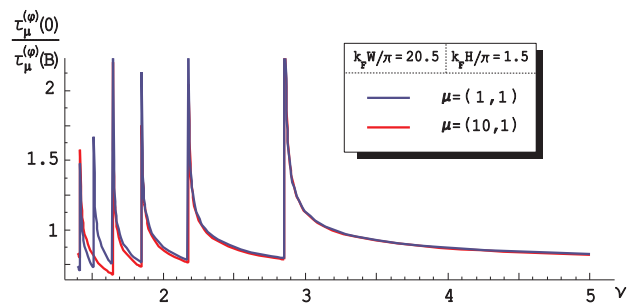}} \caption{The
dephasing rate \eqref{dephase} for two particular modes vs inverse
magnetic field. \hfill\label{fig4}}
\end{figure}

The result \eqref{deph(B)}, which is actually semiclassical, is of limited
applicability. Upon varying the magnetic field the number of extended modes
changes stepwise. Therefore the majority of physical quantities are bound to
exhibit the oscillatory behaviour, which is closely related to well-known
van~Hove singularities in MDOS. In Fig.~\ref{fig4}, the dephasing rates
obtained numerically from Eq.~\eqref{dephase} for two specific modes of
the electron waveguide are shown as functions of the inverse magnetic field.
Square-root singularities manifestly develop on both of the curves. One can
also notice that scattering frequencies for different modes start to
noticeably deviate from one another only in the range of relatively strong
magnetic fields, where the number of extended modes assumes the value
comparable with unity.

\begin{figure}[!!!h]
\vspace{.5\baselineskip} \centering\setcaptionmargin{1em}
{\includegraphics[width=.95\columnwidth]{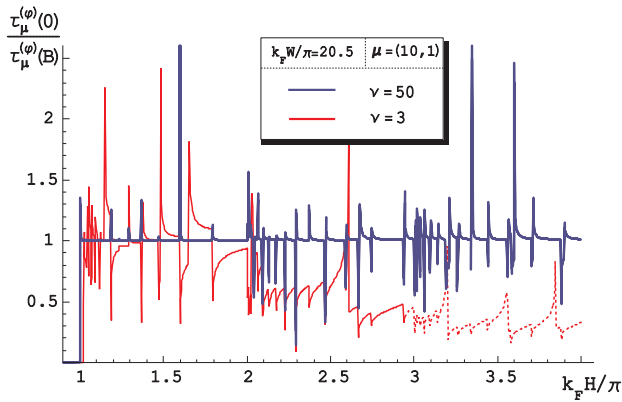}} \caption{The
dephasing rate vs the quantum waveguide thickness at different
strengths of in-plane magnetic field. The broken fraction of the
lower curve falls into the range of parameters where WMS condition
\eqref{WMS-cond} is violated. \hfill\label{fig5}}
\end{figure}

Besides the magnetic-field singularities depicted in Fig.~\ref{fig4}, in
Fig.~\ref{fig5} the dephasing rate of the particular mode vs size parameters
of the quantum waveguide is presented for two distinct values of
$|\mathbf{B}|$. Here, MDOS oscillations caused by abrupt changes in the
number of conducting channels also make themselves very evident.
We are led to conclude that by means of the orbital coupling to the electrons
in a Q2D conducting system the in-plane magnetic field can take an effect
which in some sense is analogous to that of electrostatic confinement
potential. At the same time, in contrast to the magnetic-field-controlled
singularities of the dephasing rate, which are depicted in Fig.~\ref{fig4},
oscillations of truly geometrical origin are noticeably more complicated. The
distinction is caused by substantially different response of the effective
mode energy \eqref{moden-renorm} to the magnetic field, on the one hand, and
to size parameters of the confined electron system, on the other. However, it
should be noted that in both of the graphs, 4~and~5, the reduction of the
dephasing by quenched disorder is clearly visible as the magnetic field
grows. This fact can serve as the indication of increasing coherence of
electron transport in quench-disordered Q2D systems if they are subjected to
external magnetic field.

\section{Conclusion}

In this study we have demonstrated that the observed giant positive
magnetoresistance of 2D electron and hole systems subject to parallel
magnetic field can be reasonably explained in the framework of Fermi liquid
theory being applied to structures created by confining potential wells of
finite rather than zero width. The magnetic field coupling to the carrier
orbital motion which is due to finite thickness of
\emph{quasi}-two-dimensional layers, even though rather weak from
semi-classical point of view, has been proven to influence quite essentially
the \emph{collective} electron spectrum. The reduction in the number of
extended modes with a growth of the magnetic field, as seen from
Fig.~\ref{fig2}, is very significant, continuing right up to zero in
moderately strong fields, whereas individual electron trajectories in the
plane normal to the magnetic field can go far beyond the effective thickness
of the gated carrier system. The mode truncation effect of the in-plane
magnetic field is the more noticeable the larger is the aspect ratio of the
confining potential well forming the electron waveguide.

Since the number of extended modes, according to the Landauer theory,
specifies the conductance of a~bounded system, the results presented in
Figs.~\ref{fig2} and~\ref{fig3} can be directly related to the experiment. In
fact, they may be regarded as showing the conductance dependence on the
cor\-res\-ponding parameters in the case of a \emph{perfect} confining
potential well. As the perfect we mean a~waveguide-type structure in which
any mechanism of \emph{collective} scattering of properly defined carrier
modes does not exist. This actually implies that no scattering fields other
than those involved in the unperturbed quasi-particle state formation in a
particular system are taken into account. Specifically, the collective states
pertinent to the problem considered in this study are specified by the
confining potential profile. In the absence of the disorder potential the
\emph{collective} electron motion should be regarded as ballistic, even
though individual carriers do experience strong (specular) scattering at side
boundaries of the potential well.

If some random potential is involved, e.~g., impurities or the roughness of
quantum well boundaries, it should lead to \emph{stochastic} rather than
regular scattering of the primordial quasi-particles. It seems advantageous
to separate this type of scattering into two  kinds, namely, intra- and
inter-mode scattering. The former type of scattering provides renormalization
of transport parameters and also gives rise to Anderson localization of
carrier states \emph{in the direction of current}. The latter type, inelastic
in form from the viewpoint of mode theory, leads to stochastic spreading of
mode energy levels, or, in other words, to \emph{spatial} dephasing of mode
states. At first glance, it may appear that inter-mode scattering caused
exclusively by the magnetic field is bound to produce the dephasing effect
analogous to that introduced by the quenched disorder. However, the estimate
\eqref{B/imp} is obviously contradicting to this expectation. According to
the evaluation, in the absence of random potential, which ensures
probabilistic property of mode energy levels, no imaginary part must be
contained in the mode self-energy, in spite of substantial inter-mode mixing
due to the magnetic field.

Physically, this fact seems to be quite natural. Indeed, if one chooses to
model lateral confinement of a Q2D carrier system by the quadratic rather
than the rectangular potential, eigen-functions of the transverse Hamiltonian
could be obviously selected so as to completely avoid the mode coupling due
to the magnetic field. The additional random potential, though static, would
be in this case the only cause of the mode levels widening. At the same time,
the quadratic confinement possesses the same symmetry of the confined system
as the rectangular well does. Therefore, it would be difficult to
substantiate the drastic difference of the results obtained within the
framework of these two models if one is guided by general considerations
only.

Fortunately, the result \eqref{B/imp} reveals the lack (in the
asymptotic sense) of the magnetic-field-originated dephasing of
the natural carrier spectrum. Certainly, the magnetic field does
take part in the mode level spreading, yet mostly through the
dependence on this field of the number of extended modes and of
the mode density of states. This parameters essentially determine
the impurity-originated dephasing rate \eqref{dephase}, which can
thus be viewed as being produced by the magnetic-field-dependent
disorder. The idea of the ``magnetic-field-driven disorder'' was
previously suggested in Ref.~\onlinecite{bib:PBPB01}, so the
result \eqref{dephase} can be viewed as substantiating the
rationality of such an interpretation. Clearly, in order to make a
detailed comparison with experimental observations it is necessary
to derive required formulas for the magnetocondactance. This work
will be postponed for the next publication.

\begin{acknowledgments}
  This work was partially supported by the Ukrainian Academy of
  sciences, grant No. 12/04--H under the program ``Nanostructure
  systems, nanomaterials and nanotechnologies''.
\end{acknowledgments}

\appendix*

\section{Disorder averaging of the trial Green function}

After substitution of functions \eqref{psi_pm-in} into \eqref{Green-Cochi},
the trial Green function inside magnetically biased interval $\mathcal{L}$
can be represented as a sum of four packets of spatial harmonics, viz.
\begin{widetext}
\begin{equation}
\begin{aligned}
  G^{(V)}(x,x')=\mathcal{G}_1(x,x')\mathrm{e}^{i\varkappa(x-x')}
          +\mathcal{G}_2(x,x')\mathrm{e}^{-i\varkappa(x-x')}
          +\mathcal{G}_3(x,x')\mathrm{e}^{i\varkappa(x+x')}
          +\mathcal{G}_4(x,x')\mathrm{e}^{-i\varkappa(x+x')}\ .
  \label{Green-packets}
\end{aligned}
\end{equation}
Here, smooth envelope functions are given as
\begin{subequations}\label{Green_smooth}
\begin{align}
  \label{Green_smooth-1}
 & \mathcal{G}_1(x,x')=\frac{-i}{2\varkappa}\mathcal{A}(x)
    \left[\Theta_+\frac{\pi_-(x')}{\pi_-(x)}-
    \Theta_-\frac{\gamma_+(x')}{\pi_+(x)}\Gamma_-(x)\mathrm{e}^{2i\varkappa L}\right]\ , \\
  \label{Green_smooth-2}
 & \mathcal{G}_2(x,x')=\frac{-i}{2\varkappa}\mathcal{A}(x)
    \left[\Theta_-\frac{\pi_+(x')}{\pi_+(x)}-
    \Theta_+\Gamma_+(x)\frac{\gamma_-(x')}{\pi_-(x)}\mathrm{e}^{2i\varkappa L}\right]\ , \\
  \label{Green_smooth-3}
 & \mathcal{G}_3(x,x')=\frac{-1}{2\varkappa}\mathcal{A}(x)
    \mathrm{e}^{i\varkappa L}\left[\Theta_+\frac{\gamma_-(x')}{\pi_-(x)}+
    \Theta_-\frac{\pi_+(x')}{\pi_+(x)}\Gamma_-(x)\right]\ , \\
  \label{Green_smooth-4}
 & \mathcal{G}_4(x,x')=\frac{-1}{2\varkappa}\mathcal{A}(x)
    \mathrm{e}^{i\varkappa L}\left[\Theta_-\frac{\gamma_+(x')}{\pi_+(x)}+
    \Theta_+\Gamma_+(x)\frac{\pi_-(x')}{\pi_-(x)}\right]\ ,
\end{align}
\end{subequations}
\end{widetext}
where the notations are used
\begin{subequations}\label{A_G}
\begin{align}\label{A_G-A}
  &\mathcal{A}(x)=  \left[1+\Gamma_+(x)\Gamma_-(x)\mathrm{e}^{2i\varkappa
  L}\right]^{-1}\ , \\
  \label{A_G-Gamma}
  &\Gamma_{\pm}(x)=  \gamma_{\pm}(x)/\pi_{\pm}(x)\ ,\\
  &\Theta_{\pm}=\theta[\pm(x-x')]\ .
\end{align}
\end{subequations}

As regards the functions $\Gamma_{\pm}(x)$, their physical meaning is readily
deduced from Eq.~\eqref{psi_pm-in}. They represent reflection factors of
spatial harmonics $\pm \varkappa$ incident at the point~$x$ onto the layers
with end coordinates $x$ and~$\pm L/2$, respectively. This factors meet the
Riccati-type dynamic equations,
\begin{equation}\label{Riccati}
  \pm\frac{\mathrm{d}\Gamma_{\pm}(x)}{\mathrm{d}x}=
  2i\eta(x)\Gamma_{\pm}(x)-\zeta_{\pm}(x)+\zeta_{\pm}^*(x)\Gamma_{\pm}^2(x)\ ,
\end{equation}
with boundary conditions stemming from \eqref{bound_cond},
\begin{equation}\label{BC-Gamma_pm}
  \Gamma_{\pm}(\pm L/2)=\EuScript{R}^{(B)}\ .
\end{equation}
The averaging technique for the functionals of random fields
\eqref{eff_fields} was elaborated in
Refs.~\onlinecite{bib:Tar00,bib:MakTar01,bib:FreiTar01}. Here we only briefly
indicate the main peculiarities of dealing with functionals of such a sort
and present the result of the function \eqref{Green-packets} averaging.

Having regard to correlation relations~\eqref{imp_corr} it was proven
\cite{bib:Tar00,bib:MakTar01,bib:FreiTar01} that binary correlation functions
of the effective random fields \eqref{eff_fields} under WIS conditions can be
cast to the form
\begin{subequations}\label{eff_fields-corr}
\begin{align}
  \label{eff_fields-corr-eta}
  & \big<\eta(x)\eta(x')\big>=\frac{1}{L_f^{(V)}}F_l(x-x')\ ,\\
  \label{eff_fields-corr-zeta}
  & \big<\zeta_{\pm}(x)\zeta_{\pm}^*(x')\big>=\frac{1}{L_b^{(V)}}F_l(x-x')\ ,
\end{align}
\end{subequations}
where $L_f$ and $L_b$ are the forward and the backward scattering lengths
given in \eqref{ext_length} for the particular mode~$\bm{\mu}$. The function
$F_l(x)$ has the form
\begin{equation}\label{F_l(x)}
  F_l(x)=\int_{-\infty}^{\infty}\frac{\mathrm{d}q}{2\pi}\mathrm{e}^{iqx}
  \frac{\sin^2(ql)}{(ql)^2}=\frac{1}{2l}\left(1-\frac{|x|}{2l}\right)
  \theta(2l-|x|)
\end{equation}
and plays the role of under-limiting $\delta$-function when averaging smooth
factors similar to the envelopes \eqref{Green_smooth}. Before averaging the
function \eqref{Green-packets} it makes sense to go over from functions
$\gamma_{\pm}(x)$, $\pi_{\pm}(x)$ and $\zeta_{\pm}(x)$ to phase-renormalized
functions
\begin{subequations}\label{phase_renorm}
\begin{align}
  \label{phase_renorm-gamma}
 \widetilde{\gamma}_{\pm}(x)&= \gamma_{\pm}(x)
 \exp\left[\pm i\int_x^{\pm L/2}\mathrm{d}x_1\eta(x_1)\right]\ , \\
  \label{phase_renorm-pi}
 \widetilde{\pi}_{\pm}(x)&= \pi_{\pm}(x)
 \exp\left[\mp i\int_x^{\pm L/2}\mathrm{d}x_1\eta(x_1)\right]\ , \\
 \label{phase_renorm-zeta}
 \widetilde{\zeta}_{\pm}(x)&= \zeta_{\pm}(x)
 \exp\left[\pm 2i\int_x^{\pm L/2}\mathrm{d}x_1\eta(x_1)\right]\ ,
\end{align}
\end{subequations}
which enables one to remove the forward-scattering random field $\eta(x)$
from all dynamic equations and to separate it out in the form of exponential
factors. In particular, note then that correlation relation
\eqref{eff_fields-corr-zeta} remains unchanged after renormalization
\eqref{phase_renorm-zeta}.

One can easily reveal that in view of short-range correlation of random
functions \eqref{eff_fields} and due to the causal nature of functionals
being averaged, the averaging of functionals with different sign indices in
\eqref{Green_smooth} can be done separately. By averaging the equation
\begin{equation}\label{renormGamma}
  \pm\frac{\mathrm{d}\widetilde{\Gamma}_{\pm}(x)}{\mathrm{d}x}=
  -\widetilde{\zeta}_{\pm}(x)+
  \widetilde{\zeta}_{\pm}^*(x)\widetilde{\Gamma}_{\pm}^2(x)
\end{equation}
using the Furutsu-Novikov formula for gaussian random process
\cite{bib:Klyats86} we obtain
\begin{equation}\label{AvGamma_pm-sol}
  \langle\widetilde{\Gamma}_{\pm}(x)\rangle=\EuScript{R}^{(B)}
  \exp\left[-\frac{1}{L_b}\left(\frac{L}{2}\mp x\right)\right]\ .
\end{equation}
In view of smallness of the reflection coefficient $\EuScript{R}^{(B)}$ this
allows one, when averaging \eqref{Green-packets}, to retain in
\eqref{Green_smooth} only the terms which do not contain factors
$\widetilde{\Gamma}_{\pm}(x)$ and $\widetilde{\gamma}_{\pm}(x)$.

In order to average the ratio
$\widetilde{\pi}_{\pm}(x')/\widetilde{\pi}_{\pm}(x)$, which is present in the
principal terms of \eqref{Green_smooth}, it is worthwhile to consider its
Fourier transform over $x'$ which, in view of the presence of
$\Theta$-functions in \eqref{Green_smooth}, takes the form
\vspace{.7cm}%
\begin{widetext}
\begin{equation}\label{Fourier-1}
  \Phi^{(\pm)}(x,q)=\pm\int_x^{\pm L/2}\mathrm{d}x_1
  \frac{\widetilde{\pi}_{\pm}(x_1)}{\widetilde{\pi}_{\pm}(x)}
  \exp\left[-iq(x-x_1)
  +i\varkappa|x-x_1|
  \pm i\int_{x_1}^x\mathrm{d}x_2\eta(x_2)\right]\ ,
\end{equation}
where forward-scattering random field $\eta(x)$ is already singled out. The
averaging over this field yields
\begin{equation}\label{Av_eta}
  \Big<\exp\left[\pm i\int_{x_1}^x\mathrm{d}x_2\eta(x_2)\right]\Big>_{\eta}=
  \exp\left(-\frac{|x-x_1|}{2L_f}\right)\ ,
\end{equation}
and the function \eqref{Fourier-1}, averaged beforehand over $\eta(x)$, is
found to obey the equation
\begin{equation}\label{Phi-eq}
  \mp\frac{\mathrm{d}\big<\widetilde{\Phi}^{(\pm)}(x,q)\big>_{\eta}}{\mathrm{d}x}=
  1-\left(\frac{1}{2L_f}-i\varkappa\mp iq\right)\big<\widetilde{\Phi}^{(\pm)}(x,q)\big>_{\eta}
  -\widetilde{\zeta}_{\pm}^*(x)
  \widetilde{\Gamma}_{\pm}(x)\big<\widetilde{\Phi}^{(\pm)}(x,q)\big>_{\eta} \ ,
\end{equation}
which is to be solved along with Eq.~\eqref{renormGamma}. By averaging
\eqref{Phi-eq} over the effective back-scattering field
$\widetilde{\zeta}_{\pm}(x)$ we arrive at the dynamic equations
\begin{equation}\label{Phi_aver-eq}
  \mp\frac{\mathrm{d}\big<\widetilde{\Phi}^{(\pm)}(x,q)\big>}{\mathrm{d}x}=
  1-\left[\frac{1}{2}\left(\frac{1}{L_f}+\frac{1}{L_b}\right)-
  i\varkappa\mp iq\right]\big<\widetilde{\Phi}^{(\pm)}(x,q)\big>\
\end{equation}
with obvious ``initial'' conditions $\big<\widetilde{\Phi}^{(\pm)}(\pm
L/2,q)\big>=0$. The solution to Eq.~\eqref{Phi_aver-eq} has the form
\begin{equation}\label{Phi-sol}
  \big<\widetilde{\Phi}^{(\pm)}(x,q)\big>=
  \left[\frac{1}{2}\left(\frac{1}{L_f}+\frac{1}{L_b}\right)-
  i\varkappa\mp iq\right]^{-1}
  \Bigg[1-\exp\left\{-\left[\frac{1}{2}\left(\frac{1}{L_f}+\frac{1}{L_b}\right)-
  i\varkappa\mp iq\right]\left(\frac{L}{2}\mp x\right)\right\}\Bigg]\ ,
\end{equation}
finally yielding
\begin{equation}\label{Green_1-2terms}
  \big<\mathcal{G}_1(x,x')\big>\mathrm{e}^{i\varkappa(x-x')}+
  \big<\mathcal{G}_2(x,x')\big>\mathrm{e}^{-i\varkappa(x-x')}\approx
  \frac{-i}{2\varkappa}
  \exp\left\{\left[\mathrm{i}\varkappa-\frac{1}{2}
  \left(\frac{1}{L_f}+
  \frac{1}{L_b}\right)\right]|x-x'|\right\} \ .
\end{equation}
\end{widetext}

The envelopes \eqref{Green_smooth-3} and \eqref{Green_smooth-4} can be
averaged in the same manner. Because both of them are proportional to
reflection coefficient $\gamma_{\pm}(x)$, they prove to be relatively small
in the parameter~\eqref{WMS-cond} and can thus be omitted, leaving the
result\eqref{Green_1-2terms} as the main approximation for the
impurity-averaged trial Green function.



\begin{thebibliography}{cc}
\bibitem{rem1}
In Ref.~\onlinecite{bib:PBPB02}, experimental results consistent with this
observation were obtained.
\end{thebibliography}
\end{document}